# Magnetostriction in microwave synthesized $La_{0.5}Ba_{0.5}CoO_3$

M. Manikandan, A. Ghosh, and R. Mahendiran[*]

*Department of Physics, 2 Science Drive 3, National University of Singapore,*

*Singapore 117551, Republic of Singapore*

## ABSTRACT

A single phase polycrystalline $La_{0.5}Ba_{0.5}CoO_{3-d}$ sample possessing cubic structure (space group $Pm\bar{3}m$) was synthesized by microwave irradiation within 20 minutes of processing time and its structural, magnetic, electrical, and magnetostrictive properties were investigated. While the temperature dependence of field-cooled magnetization ($M$) in a field of $H = 0.5$ kOe indicates onset of ferromagnetic transition at $T_C = 177$ K, irreversibility between the zero field-cooled and field cooled $M(T)$ persists even at $H = 3$ kOe. $M(H)$ at 10 K does not saturate at the maximum available field and has much smaller value (0.87 $\mu_B$/Co in a field of 50 kOe) than 1.9 $\mu_B$/Co expected for spin only contribution from intermediate $Co^{3+}$ and $Co^{4+}$ spins. Resistivity shows insulating behaviour down to 10 K and only a small magnetoresistance (~ 2% for $H = 70$ kOe) occurs around $T_C$. All these results suggest a magnetically heterogeneous ground state with weakly interacting ferromagnetic clusters coexisting with a non-ferromagnetic phase. The length of the sample expands in the direction of the applied magnetic field (positive magnetostriction) and does not show saturation even at 50 kOe. The magnetostriction has a maximum value ($\lambda_{par} = 252 \times 10^{-6}$) at 10 K and it decreases with increasing temperature. The smaller value of $\lambda_{par}$ compared to the available data on $La_{0.5}Sr_{0.5}CoO_3$ ($\lambda_{par} = 900 \times 10^{-6}$) suggests that non-ferromagnetic matrix is most likely antiferromagnetic and it restrains the field-induced expansion of ferromagnetic clusters in the microwave synthesized $La_{0.5}Ba_{0.5}CoO_{3-d}$ sample.

---
[*] Author for correspondence : R. Mahendiran (email: phyrm@nus.edu.sg)

# 1. Introduction

The family of hole-doped cobaltites with the general formula $R^{3+}_{1-x}A^{2+}_{x}CoO_3$, where $R$ is a rare earth and $A$ is an alkaline earth cation, has been a topic of interest for many years because of various interesting physical phenomena exhibited by them such as the rare occurrence of ferromagnetism and metallicity among $3d$ perovskite oxides other than manganites[1,2,3], spin-state transition of Co ions[4], spin polarons for low doping[5,6,7,8], giant magnetoresistance[9,10,11] etc. The undoped LaCoO$_3$ containing only Co$^{3+}$ ions is a non-magnetic insulator and it transforms into a ferromagnetic metal upon doping holes (Co$^{4+}$:d$^5$) into Co$^{3+}$:d$^6$ matrix for $x \approx 0.25$ in La$_{1-x}$A$_x$CoO$_3$ (A = Sr, Ba, Ca) and the ferromagnetic transition temperature increases with $x$ up to $x = 1.0$ [2,12,13]. While Co$^{3+}$ ion is in the low-spin state (t$_{2g}^6$e$_g^0$, S = 0) at liquid helium temperature in the undoped LaCoO$_3$, intermediate- spin state (t$_{2g}^5$e$_g^1$, S = 1) is stabilized in ferromagnetic metallic compositions of La$_{1-x}$Sr$_x$CoO$_3$. While the structural, electrical and magnetic properties of various combinations of $R$ and $A$ cations have been studied extensively[14], only few studies are available on the interplay between magnetism and elastic properties of these oxides. Ibarra et al.[15] reported giant anisotropic magnetostriction ($\lambda_t \sim 10^{-3}$)- a large difference between the strain measured parallel and perpendicular to the direction of applied field- in ferromagnetic metallic compositions ($x = 0.3$ and 0.5) of La$_{1-x}$Sr$_x$CoO$_3$ series, which is rare among 3d perovskite oxides. They attributed the origin of giant anisotropic magnetostriction to field induced spin state transition of Co$^{3+}$ ions from low spin (t$_{2g}^6$e$_g^0$, S = 0) to Jahn-Teller active intermediate spin (t$_{2g}^5$e$_g^1$, S = 1). Kundys et al.[16] investigated magnetoelastic effect in La$_{0.7}$Sr$_{0.3}$CoO$_3$ thin films and suggested two mechanisms for the magnetostriction: magnetic field induced increase in volume of ferromagnetic clusters and magnetic field induced spin-state transition of Co$^{3+}$ ions. Contrary to the giant anisotropic magnetostriction effect found in Sr-doped LaCoO$_3$, Rotter et al.[17] found anomalous volume expansion in the parent compound LaCoO$_3$ which was attributed to low-spin to high-spin (t$_{2g}^4$e$_g^2$) state transition of Co$^{3+}$ ions under external magnetic fields. However, a magnetic field H ≥ 60 T was needed to induce such a positive magnetovolume effect in LaCoO$_3$ in contrast to few Tesla magnetic fields needed to induce anisotropic magnetostriction in the ferromagnetic compositions. Interestingly, magnetic shape memory effect in twined rhombohedral

La$_{0.8}$Sr$_{0.2}$CoO$_3$ single crystal was reported recently.[18] : A magnetic field of $H$ = 6 T induces twin boundary motion at room temperature and magnetostriction even though its ferromagnetic Curie temperature is well below the room temperature. The critical field needed to induce twin-boundary motion decreases with lowering temperature.

Since Ba$^{2+}$ cation has larger average ionic radius ($r_a$ = 1.46 Å) than the Sr$^{2+}$ cation ($r_a$= 1.31 Å) and La$^{3+}$ ($r$ = 1.22 Å), it reduces the overall distortion from the ideal cubic structure and the tolerance factor approaches unity. The crystal structure of La$_{0.5}$A$_{0.5}$CoO$_3$ changes from rhombohedral for A = Sr to cubic for A = Ba and the ferromagnetic Curie temperature decreases from ~ 248 K for the Sr cation to ~ 180-190 K for the Ba cation.[19,20,21] La$_{0.5}$Ba$_{0.5}$CoO$_3$ adopts different lattice forms depending on the synthesis methods.[22, 23] The unit cell of A-site ordered LaBaCo$_2$O$_6$ has tetragonal symmetry (space group I4/mcm), in which La and Ba cations alternatively order in LaO- BaO layers along the $c$-axis of the tetragonal unit cell. The disordered La$_{0.5}$Ba$_{0.5}$CoO$_3$ has a cubic structure wherein the A-site is randomly occupied by La and Ba cations and there is also a nanoscale ordered LaBaCo$_2$O$_6$ phase.[24] Interestingly, all these variants have close Curie temperatures ($T_C$ ~ 175-190 K). According to Fauth *et al.*[17] ferromagnetism in the A-site disordered La$_{0.5}$Ba$_{0.5}$CoO$_3$ is accompanied by long-range tetragonal distortion arising from the cooperative Jahn-Teller distortion associated with the intermediate spin states of Co$^{3+}$ and Co$^{4+}$ ions. A very interesting thermal expansion behaviour was reported in Ba-doped samples by Tan *et al.*[25] who studied thermal dependence of the lattice parameters in La$_{0.5}$Ba$_{0.5}$CoO$_{3-x}$ system with controlled oxygen deficiency ($x$ ~ 0.04- 0.23) using high resolution neutron and synchrotron X-ray diffraction methods. All the oxygen deficient samples were found to be cubic at room temperature but oxygen deficiency was found to promote G-type antiferromagnetic ordering at low temperatures. The Neel temperature ($T_N$) increased with the oxygen deficiency ($T_N$ = 164 K for $x$ = 0.1 and 390 K for $x$ = 0.20). While the lattice parameter for $x$ = 0.04 and 0.06 samples decreased smoothly with decreasing temperature below room temperature, negative thermal expansion was found below $T_N$ in $x$ = 0.10 – 0.18 and the samples with still higher oxygen deficiency showed zero thermal expansion over a certain temperature range, mimicking the behaviour of INVAR alloys. Neutron diffraction under external magnetic fields recorded for $x$ = 0.12 at 100 K indicates a transition from antiferromagnetic (high volume phase) to ferromagnetic phase (low volume) with increasing strength of magnetic field although cubic symmetry was maintained from 300 K to 2 K. These results motivated us to study the magnetic field dependence of macroscopic strain in bulk La$_{0.5}$Ba$_{0.5}$CoO$_{3-d}$.

In this paper, we report synthesis of $La_{0.5}Ba_{0.5}CoO_{3-d}$ by microwave irradiation and its magnetic, electrical and magnetostrictive properties. In microwave assisted synthesis, electrical dipoles in a sample oscillates in response to microwave electrical field and the sample absorb a maximum microwave power when the electrical dipoles are driven to resonance.[26,27,28] The absorbed power is dissipated internally as volumetric heating and large dielectric loss promotes heating. Thus, heat is internally generated in a sample in microwave heating in contrast to transfer of heat by thermal convection from heating element to the sample in an electrical furnace. Due to rapid heating, the final product can be obtained within few tens of seconds to tens of minutes compared to several days needed to obtain the final product using a conventional electrical furnace.[26] Gutierrez Sijas *et al.*[29] reported microwave synthesis of $RCoO_3$ (R = La-Dy) and studied their magnetic susceptibilities. However, neither microwave synthesis of $La_{0.5}Ba_{0.5}CoO_3$ and nor its magnetostrictive property has been reported so far.

## 2. Experimental details

A stoichiometric mixture of dehydrated $La_2O_3$, $BaCO_3$ and $Co_3O_4$ was ground well with an agate mortar and pestle. The powder was initially decarbonated at 1200 °C for 10 minutes in a muffle microwave furnace (Milestone PYRO mode MA 194-003) operating at 2.45 GHz with 1400 W power. After regrinding, the decarbonated powder was pressed into a circular pellet and irradiated with a microwave power of 1600 W. The temperature was set to reach 1200°C in 20 min under microwave irradiation and then maintained at 1200°C for 20 min. Then, the microwave power was switched off and the pellet was allowed to cool to room temperature in 4 hours (300ºC/hr). A portion of the irradiated pellet was crushed into powder and analysed for phase structural characteristics using X-ray diffraction (XRD) with Cu-$K\alpha_1$ (1.5406 Å) radiation. A pinch of crushed powder was spread over a carbon tape stuck on the aluminium grid and used to analyse the microstructure in field emission scanning electron microscopy (FE-SEM, JEOL JSM-6700F). Magnetization was measured using a vibrating sample magnetometer probe that can be inserted in a physical property measurements system (PPMS). Magnetostriction along the direction of the applied *dc* magnetic field was measured using a capacitance dilatometer probe that is inserted in the PPMS. A polished cube shaped sample of size 2 x 2 x 2 mm filled the space between two circular capacitive electrodes. The capacitance change of the dilatometer was measured using a high resolution capacitance bridge (Andeen Hagerling, model AH2500A) in the temperature range 300 K to 10 K. For the isothermal field sweep measurements, the field sweep rate was fixed to 60 Oe/sec for both the magnetization and magnetostriction. Magnetostriction is obtained as $\lambda_{par} = [L(H)-L(0)/L(0)] \times 10^{-6}$ where $L(H)$ and $L(0)$ are the lengths of the sample at a fixed temperature in a magnetic field $H$ and zero field respectively, measured in the direction of the applied dc magnetic field.

## 3. Results and Discussion

Fig. 1 shows the room temperature X-ray diffraction patterns of the MW synthesized $La_{0.5}Ba_{0.5}CoO_{3-d}$. The well-defined diffracted peaks disclose excellent crystallization of the sample. Bragg's reflections are indexed with cubic symmetry (space group of $Pm\bar{3}m$). The refined cell parameters of $a = 3.886(2)$ Å and volume ($V \approx 59$ Å$^3$) are close to the cell parameters reported for a disordered phase of stoichiometric $La_{0.5}Ba_{0.5}CoO_3$[5,6]. In $Pm\bar{3}m$ crystallographic structure, both La and Ba ions are statistically distributed at the body centre site, whereas the Co ions are at the centre of each octahedra formed by its six neighbouring oxygen ions. FE SEM micrograph illustrated in the inset of figure suggest average grain size of 3 micron.

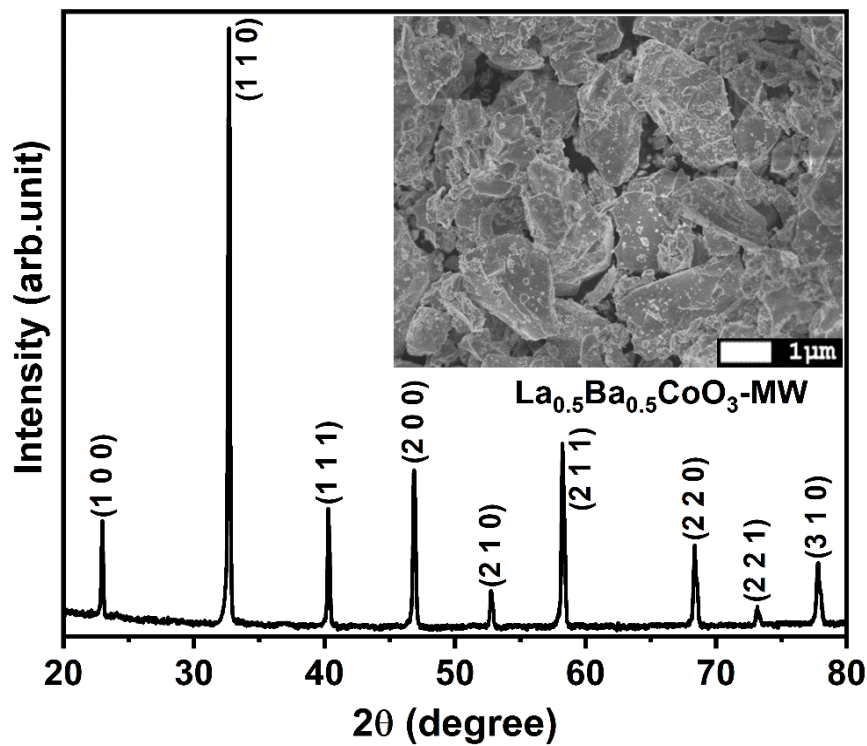

**FIG. 1**. Powder X-ray diffraction pattern of the microwave synthesized $La_{0.5}Ba_{0.5}CoO_{3-d}$. The reflections lines are fitted using XRDA software. Inset shows field emission microscope image of distribution of grains.

Figure 2(a) shows the temperature dependence of magnetization ($M$) of La$_{0.5}$Ba$_{0.5}$CoO$_{3-d}$ measured upon warming from 10 K after zero-field cooling (ZFC) and field-cooling (FC) to 10 K for three values of the magnetic field ($H$ = 0.5, 1 and 3 kOe). The sample undergoes a paramagnetic to ferromagnetic transition at $T_C$ = 177 K. The $T_C$ is obtained from the inflection point of $dM/dT$ for $H$ = 0.5 kOe in the FC mode (shown in the inset). The ZFC curve for each field increases as the temperature is increased from 10 K and goes through a maximum value at the temperature $T_m$, which is much below $T_C$. With increasing strength of the magnetic field, $T_m$ shifts towards low temperatures ($T_m$ = 128, 108, 71 K for H = 0.5, 1, and 3 kOe, respectively) and the difference between ZFC and FC curves also decrease. The irreversibility of the magnetization together with low value of magnetization at 10 K for $H$ = 50 kOe (shown later in figure 3) suggest that the ferromagnetism is short-range in this compound.

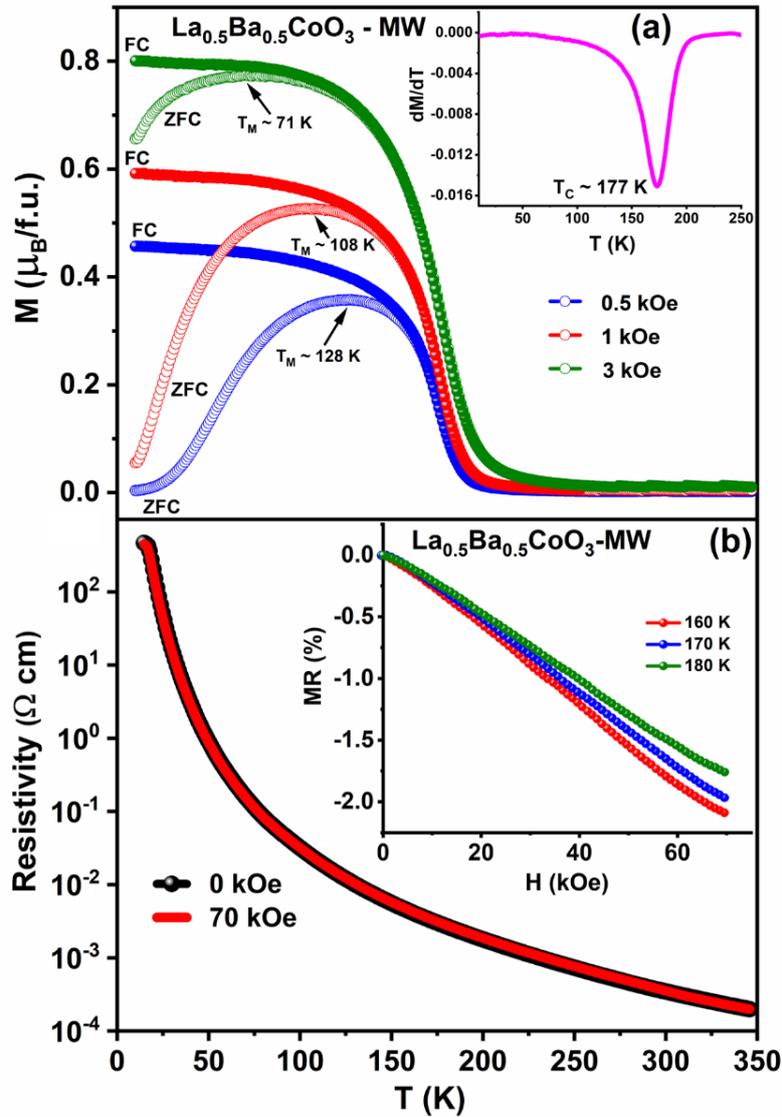

**FIG. 2.** (a) Temperature dependence of magnetization while warming after zero-field cooling (ZFC, open symbols) and field cooling (FC, closed symbols) at three different magnetic fields (H = 0.5, 1 and 3 kOe) in the microwave synthesized $La_{0.5}Ba_{0.5}CoO_{3-d}$. The inset shows $dM/dT$ versus temperature for $H = 0.5$ kOe in the ZFC mode. $T_C$ is the ferromagnetic Curie temperature. (b) Temperature dependence of *dc* resistivity in zero and under $H = 70$ kOe. The inset shows the field dependence of magnetoresistance at three temperatures around $T_C$.

The temperature dependence of the resistivity ($\rho$) is shown in Fig. 2(b) for both $H = 0$ and 70 kOe from 350 K to 10 K. $\rho(T)$ shows insulating behavior: ρ increases by seven orders of magnitude, from ~ 1.9 x $10^{-3}$ Ω cm at 350 K to ~ 4.5 x $10^4$ Ω cm at 20 K. The applied magnetic field of 70 kOe does not destabilize the insulating behavior. Since the magnetoresistance appears to be small in the temperature sweep mode, we have measured magnetoresistance in the field sweep mode at three selected temperatures for clarity. The inset shows the magnetic field dependence of the magnetoresistance around $T_C$ at $T = 160$, 170 and 180 K. The magnetoresistance increases with the magnetic field without saturation and the maximum value is small (~ -2% for 70 kOe at 180 K). The magnetoresistance is negative and its magnitude increases with the magnetic field without saturation. The maximum magnitude of magnetoresistance is only 2% for 70 kOe at 180 K which is smaller ~ 6% than found in the Sr analogue ($La_{0.5}Sr_{0.5}CoO_3$)[15]

We would like to remind the readers that according to available reports in literature, $\rho(T)$ of $La_{0.5}Ba_{0.5}CoO_{3-d}$ in zero magnetic field exhibits different trends depending on the synthesis conditions. Nakajima *et al.*[22] reported that $\rho(T)$ of A-site disordered $La_{0.5}Ba_{0.5}CoO_3$ ($T_C = 190$ K) sample synthesized by solid state reaction at 1300 °C and annealed in oxygen changed from metallic like to semiconducting like while cooling below 140 K. Fauth *et al.*[19] found semiconducting like behaviour below 120 K and metallic behaviour above. They attributed the semiconducting like behaviour to gradual ordering of the $d_{3z^2-r^2}$ orbitals which blocked a channel for electron hopping. $\rho(T)$ showed metallic like behaviour from 300 K to 10 K in the sample of Rautuma *et al.*,[23]. However, Tryonchuk *et al.*[30] found only a weak bump around $T_C$ in zero field resistivity in a $La_{0.5}Ba_{0.5}CoO_3$ sample which was cooled at a rate of 100°C/hour after sintering at 1400°C for 24 h in air. However, the resistivity of that sample increased by a factor of three between 300 K and 10 K unlike six orders of magnitude change in our present sample. The magnitude and the temperature dependence of the resistivity could be influenced by the presence of both grain boundaries and oxygen non-stoichiometry. For comparison, we have also synthesized $La_{0.7}Ba_{0.3}CoO_3$ by the MW irradiation method and measured its electrical and magnetic properties (not shown here). $La_{0.7}Ba_{0.3}CoO_{3-d}$ showed metallic behavior in the temperature range T = 300 K-125 K and higher magnetization value (M= 1.7 $\mu_B$/Co at H = 50 kOe and T = 10 K) compared to $La_{0.5}Ba_{0.5}CoO_{3-d}$. Since grain boundary resistance is not expected to be different between these samples, the insulating-like behavior of $La_{0.5}Ba_{0.5}CoO_3$ is most likely due to oxygen deficiency. It is known the oxygen deficiency in the cobaltite system increase with increasing with divalent cation content or decreasing size of the rare earth ion. In addition, we have also synthesized $La_{0.5}Ba_{0.5}CoO_{3-d}$ by solid state reaction method. After sintering the sample at 1200° C for 24 hr, the sample was cooled to room temperature at a rate of 60 C/hr. The resistivity showed a small peak around $T_C$ (= 177 K) and the resistivity at 10 K was $10^{-3}$ohm cm, which is five orders of magnitude lower than the value exhibited by the MW synthesised $La_{0.5}Ba_{0.5}CoO_{3-d}$, which also indirectly support that the oxygen deficiency is more in the MW synthesised sample.

Figure 3 illustrates the magnetic field dependence of magnetization *M(H)* of $La_{0.5}Ba_{0.5}CoO_{3-d}$ on left *y*-axis and magnetostriction ($\lambda_{par}$) on right *y*-axis at 10 K. *M(H)* was measured after zero field cooling the sample. *M(H)* exhibits a wide hysteresis loop which is opened up to ± 10 kOe and a large coercive field ($H_C$ = 1.86 kOe). *M(H)* does not show saturation up to *H* = 50 kOe, where it reaches 0.83 $\mu_B$/Co ion, which is much lower than 1.75-1.9 $\mu_B$/Co reported at the same field by other authors.[17,19,20,29] On the other hand, Tryounchuk et al.[30] noted that the value of magnetization at 10 K for *H* = 50 kOe decreased from 1.75 $\mu_B$/Co for the sample cooled at a rate of 100°C/hr (in air from 1200°C to 300°C) to 1 $\mu_B$/Co in the sample which was cooled at a rate of 300 °C//hr. In our experiment also, the cooling rate was ~ 300 °C//hr, and thus there could be significant oxygen deficiency, which will decrease the $Co^{4+}$ content.

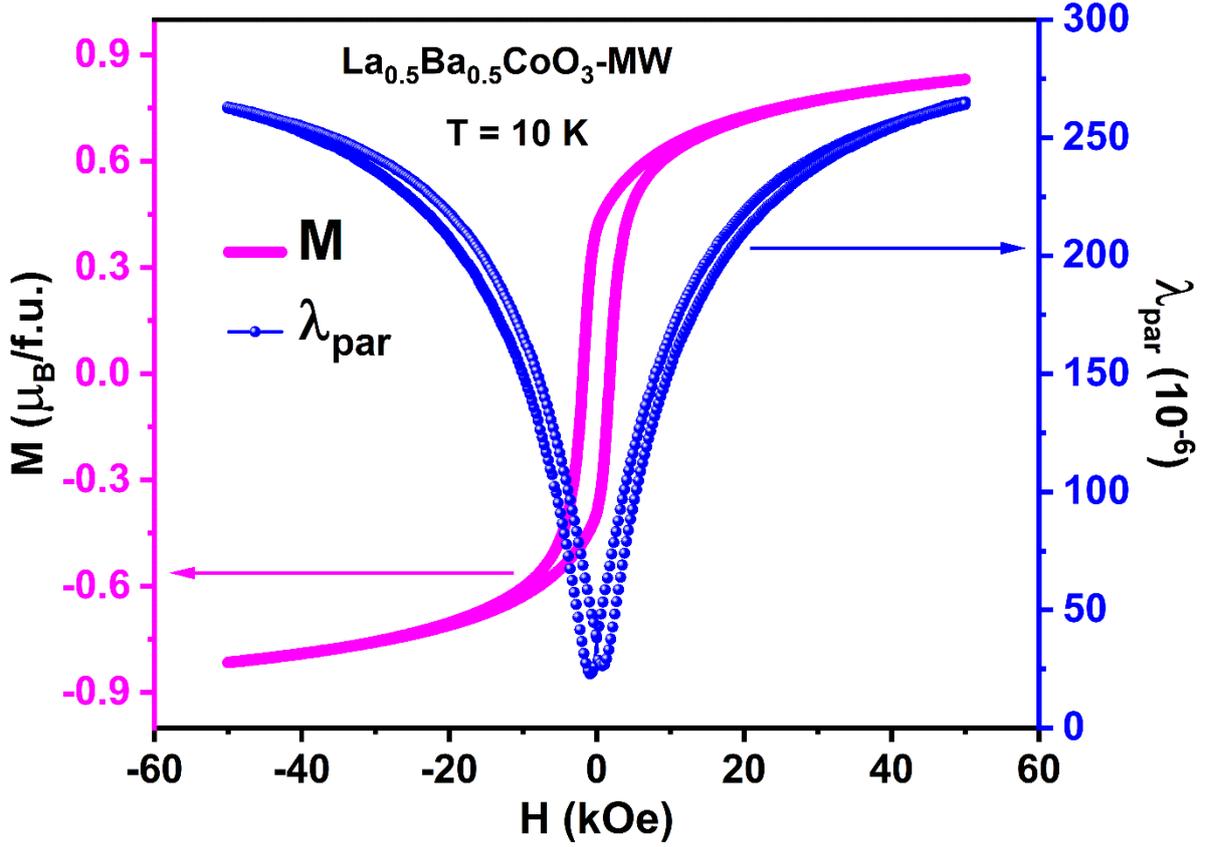

**FIG. 3.** Hysteresis loops of magnetization (*M*) on the left y-scale and parallel magnetostriction ($\lambda_{par}$) on the right y-scale at 10 K.

The parallel magnetostriction $\lambda_{par}$ is positive, *i.e.*, the length of the sample elongates along the direction of the applied magnetic field. Though $\lambda_{par}$ curve bend above 30 kOe, it does not show saturation up to the maximum field of 50 kOe. $\lambda_{par}$ reaches a maximum value of 265 x 10$^{-6}$ at 50 kOe. $\lambda_{par}$ also exhibits pronounced hysteresis up to $H \sim \pm 30$ kOe, above which curves traced while increasing and decreasing the magnetic field merge. The inset of Fig. 4 shows the field dependence of $\lambda_{par}$ vs *H* at selected temperatures. Hysteresis is visible in the isotherms up to 160 K but the field range narrows and hysteresis is negligible for 180 K and 200 K. We collect the values of $\lambda_{par}$ at the highest field from the $\lambda_{par}$ vs *H* isotherms and plot them as a function of temperature in the main panel of Fig. 4. $\lambda_{par}$ has a maximum value at 10 K, decreases with increasing temperature and becomes negligible above $T_C$. The maximum value of $\lambda_{par}$ in $La_{0.5}Ba_{0.5}CoO_{3-d}$ is much smaller than the value ($\lambda_{par}$ = 900 x $10^{-6}$) reported for the ferromagnetic $La_{0.5}Sr_{0.5}CoO_{3-d}$ at 25 K, which is expected to increase slightly at 10 K.[13] To explain the unusual high value of magnetostriction found in $La_{0.5}Sr_{0.5}CoO_{3-d}$, Ibarra *et al*.[13] suggested orbital instability of $Co^{3+}$ ions following the field induced spin-state transition from non-degenerate low-spin state ($t_{2g}^6 e_g^0$, $S = 0$) to intermediate spin state ($t_{2g}^5 e_g^1$, $S = 1$). With increasing magnetic field, the triply degenerate $t_{2g}$ energy levels split into an orbital singlet ($d_{xy}$) with zero orbital moment as a ground state and an orbital doublet ($d_{yz}$, $d_{xz}$) with the unquenched orbital moment as a first excited state. The non-zero angular momentum of the orbital doublet couples to the spin angular momentum to create intra-atomic spin-orbit coupling. As the spins rotate, orbital also distorts which causes a large magnetostriction along the field direction in $La_{0.5}Sr_{0.5}CoO_{3-d}$. Based on the non-saturation of *M*(*H*) at 10 K with a small value magnetization at 50 kOe, magnetic irreversibility between ZFC and FC-M(T) curves, and insulating behaviour of the resistivity, we can exclude long range ferromagnetism in the MW synthesized $La_{0.5}Ba_{0.5}CoO_{3-d}$. Rapid cooling in MW synthesize promoted oxygen deficiency in this compound and inhomogeneous ferromagnetic state is created. Ferromagnetism sets in small regions (clusters) where $Co^{3+}$ and $Co^{4+}$ are in the intermediate spin state. These ferromagnetic clusters are dispersed in non-ferromagnetic phase and they arenon-percolating. The non-ferromagnetic phase is most likely to be antiferromagnetic rather than paramagnetic. In the antiferromagnetic phase, $Co^{3+}$ and $Co^{4+}$ ions are in high spin states which favour antiferromagnetic interaction among same valence Co ions and between 4+ and 3+ valance Co ions unlike the intermediate spin state they adopt in the metallic phase. Antiferromagnetic spin ordering detected by neutron diffraction and increase in Neel temperature with increasing oxygen deficiency[24] in $La_{0.5}Ba_{0.5}CoO_{3-x}$ support the view that the non-ferromagnetic phase is likely to be antiferromagnetic. The Neel temperature for *x* = 0.1 was found to be 164 K.[24] It is possible that antiferromagnetic ordering happens just below the ferromagnetic transition in our sample. In the background of strong signal from ferromagnetic clusters, antiferromagnetic transition is masked in the *M*(*T*) curve. Based on the absence of non-divergence in the field dependence of the third order non-linear magnetic susceptibility, Kumar *et al*.[31] proposed that their $La_{0.5}Ba_{0.5}CoO_{3-d}$ sample consisted of ferromagnetic and antiferromagnetic clusters. Their sample showed higher magnetization (1.9 $\mu_B$/Co for H = 50 kOe at 10 K) than our sample. On the other hand, *M*(*T*) of the Ba-rich composition $La_{0.45}Ba_{0.55}CoO_{2.85}$ exhibited a pronounced peak at 100 K both in the ZFC and FC modes and neutron diffraction revealed G-type antiferromagnetic ordering.[32] Hence, presence of antiferromagnetic phase in our compound cannot be neglected. The lower value of magnetostriction in $La_{0.5}Ba_{0.5}CoO_{3-d}$ compared to $La_{0.5}Sr_{0.5}CoO_3$ is possibly due to smaller volume fraction of the ferromagnetic phase in the former sample. The antiferromagnetic phase does not change into ferromagnetic phase under 50 kOe but hinders the expansion of ferromagnetic clusters under a magnetic field.

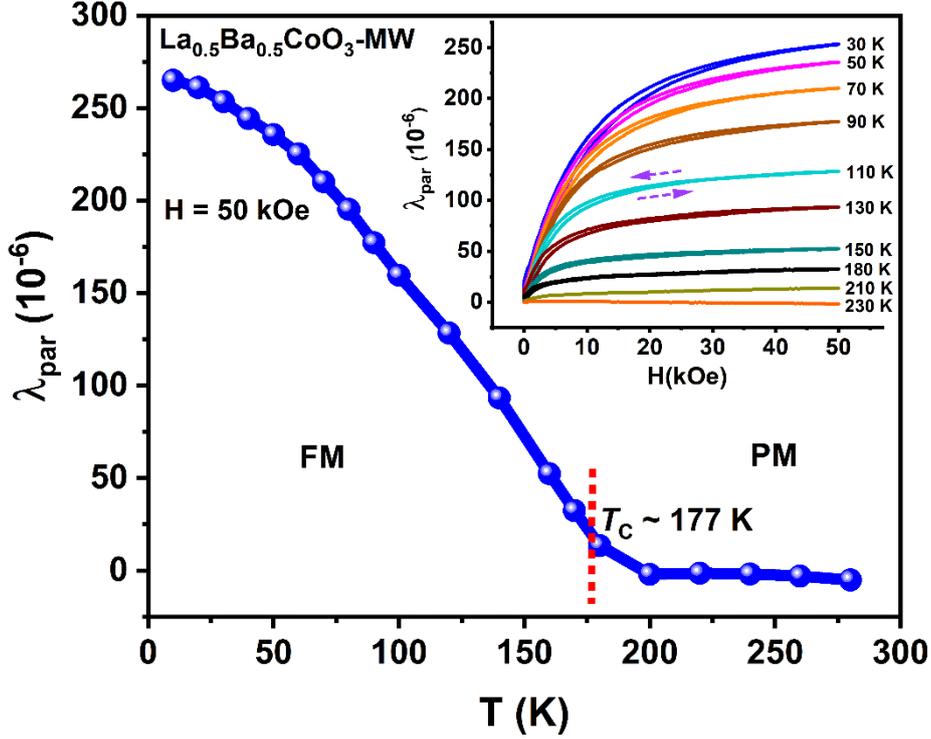

**FIG. 4.** Temperature dependence of parallel magnetostriction ($\lambda_{par}$) collected at $H = 50$ kOe from $\lambda_{par}(H)$ isotherms. Inset shows the field dependence of $\lambda_{par}$ at selected temperatures.

## 4. Summary

In summary, we have synthesized $La_{0.5}Ba_{0.5}CoO_{3-d}$ by microwave irradiation method in short time. Although the sample exhibits ferromagnetic transition at 177 K similar to the available data on solid state synthesized and oxygen annealed $La_{0.5}B_{0.5}CoO_3$ sample, the MW synthesized sample shows much reduced magnetization (~0.84$\mu_B$/Co) compared to ~1.9 $\mu_B$/Co reported in oxygen stoichiometric compound. It is suggested that non percolating ferromagnetic clusters coexist with antiferromagnetic phase below the Curie temperature of ferromagnetic clusters. Magnetostriction is positive and a maximum value (265 x 10$^{-6}$) is obtained at 10 K. Future work need to be focused on measuring both parallel and perpendicular magnetostrictions to calculated anisotropic magnetostriction and volume magnetostriction.

Also, the dependence of the magnitude of magnetostriction on the oxygen content needs to be systematically investigated.

**Acknowledgements**

R. M. acknowledges the Ministry of Education, Singapore, for supporting this work (Grants numbers: R1444-000-422-114 and R144-000-428-114).

**DATA AVAILABILITY**

The data that support the findings of this study are available from the corresponding author upon reasonable request.